\documentclass[
    ,final            
  ]
  {aipproc}

\layoutstyle{6x9}
\usepackage{graphicx}
\usepackage{amssymb}
\usepackage{amsmath}
\usepackage{units}
\usepackage{url}
\usepackage{wrapfig}
\usepackage[footnotesize, bf]{caption}
\renewcommand{\figurename}{FIGURE}

\makeatletter
\def\url@leostyle{%
  \@ifundefined{selectfont}{\def\UrlFont{\sf}}{\def\UrlFont{\small\ttfamily}}}
\makeatother
\graphicspath{{Figs/}}
\DeclareGraphicsExtensions{.eps,.ps,.eps.gz,.ps.gz}
\newcommand{\pslash}{p\llap{/\kern+0.1em}}

\newcommand{\xifb}{\ensuremath{\:\textrm{fb}^{-1}}}

\newcommand{\xGeV}{\ensuremath{\:\textrm{Ge\kern -0.1em V}}}
\newcommand{\GeV}{\ensuremath{\textrm{Ge\kern -0.1em V}}}
\newcommand{\xGeVs}{\ensuremath{\:\textrm{Ge\kern -0.1em V^{2}}}}
\newcommand{\xMeV}{\ensuremath{\:\textrm{Me\kern -0.1em V}}}
\newcommand{\MeV}{\ensuremath{\textrm{Me\kern -0.1em V}}}
\newcommand{\KeV}{\ensuremath{\:\textrm{ke\kern -0.1em V}}}
\newcommand{\eV}{\ensuremath{\:\textrm{e\kern -0.1em V}}}

\newcommand{\PYTHIA}{\textsc{Pythia}}
\newcommand{\RAPGAP}{\textsc{Rapgap}}

\begin{document}

\title{Jets and Heavy Flavors at HERA}


\author{R.~Shehzadi {\footnotesize{(for the H1 and ZEUS Collaborations)}}}{
  address={Physikalisches Institut, Universit\"at Bonn, Germany}
}

\begin{abstract}
Recent results on jet cross sections and heavy-flavor production in photoproduction and neutral current deep inelastic $ep$ scattering from the H1 and ZEUS Collaborations are presented. The jet measurements are used to perform stringent tests of perturbative QCD, to extract precise values of the strong coupling and to constrain further the proton and photon parton distribution functions. The measurement of beauty and charm production at HERA is an important testing ground for perturbative QCD and can provide information on the structure of the proton.
\end{abstract}

\maketitle


\section{Introduction}
The $ep$ collider, HERA, operated from 1992-2007 with protons of
energy $920$ $(820)\xGeV{}$ and electrons or positrons of energy
$27.5\xGeV{}$.
By the end of the running, each of the colliding-beam experiments, H1
and ZEUS, had collected about $0.5\xifb{}$ of data.

Different kinematic variables are used to describe $ep$ interactions:
the virtuality of the exchanged boson, $Q^{2}$, the Bjorken scaling
variable, $x$, and the inelasticity, $y$.  At HERA, two kinematic
regimes are distinguished depending on $Q^{2}$: photoproduction
($\gamma p$), where $Q^{2} \approx 0\xGeV^{2}$, and deep inelastic
scattering (DIS), where $Q^{2} \gtrsim 0\xGeV^{2}$.

In the following a small selection of recent measurements of jets and
heavy quark production in $\gamma p$ and DIS at HERA is presented.
\section{Jet Physics at HERA}
At leading order (LO) in $\alpha_{s}$, jet production in neutral
current (NC) DIS, can be described via the boson-gluon fusion and QCD
Compton processes. In $\gamma p$, two processes are relevant: the direct
process, in which the photon interacts as a point-like particle, and the
resolved process, in which the photon interacts through its partonic
content. Measurements of jet production in $\gamma p$
and NC DIS provide a powerful tool for stringent tests of pQCD
calculations.

In recent analyses, jet production was measured at HERA in
different kinematic regions. In these analyses jets were defined using
the $k_{T}$ clustering algorithm. For DIS, the jet algorithm was
applied in the Breit frame, in which the photon and proton collide
head on, and for $\gamma p$, in the laboratory frame. The
measurements from the ZEUS collaboration include the dijet cross
section in $\gamma p$~\cite{abr10a} ($Q^{2} < 1\xGeV^{2},
E_{\text{T}}^{\text{jet, 1(2)}} > 21 (17)\xGeV{}$) and inclusive-jet
cross sections in $\gamma p$~\cite{abr10b} ($Q^{2} < 1\xGeV^{2},
E_{\text{T}}^{\text{jet}} > 17\xGeV{}$) and high $Q^{2}$
DIS~\cite{abr10c} ($Q^{2} > 125\xGeV^{2},
E_{\text{T,B}}^{\text{jet}} > 8\xGeV{}$). 
The H1 measurements include inclusive, 2-jet and 3-jet cross sections
as well as the ratio of 2-jet to 3-jet cross sections for low
$Q^{2}$~\cite{aar10a} ($5 < Q^{2} < 100\xGeV^{2}, 5 < p_{\text{T,
    jet}} < 80\xGeV{}$) and high $Q^{2}$~\cite{aar10b} ($150 < Q^{2}
< 15000\xGeV^{2}, 7 < p_{\text{T, jet}} < 50\xGeV{}$). For the latter,
the jet cross sections were normalized to the inclusive DIS cross
section, which significantly reduces the experimental and theoretical
errors.

\begin{figure}[h]
  \hspace*{-0.4cm}\includegraphics[height = 0.4\textwidth, width=0.39\textwidth, bb = 0 0 550 535, clip = true]{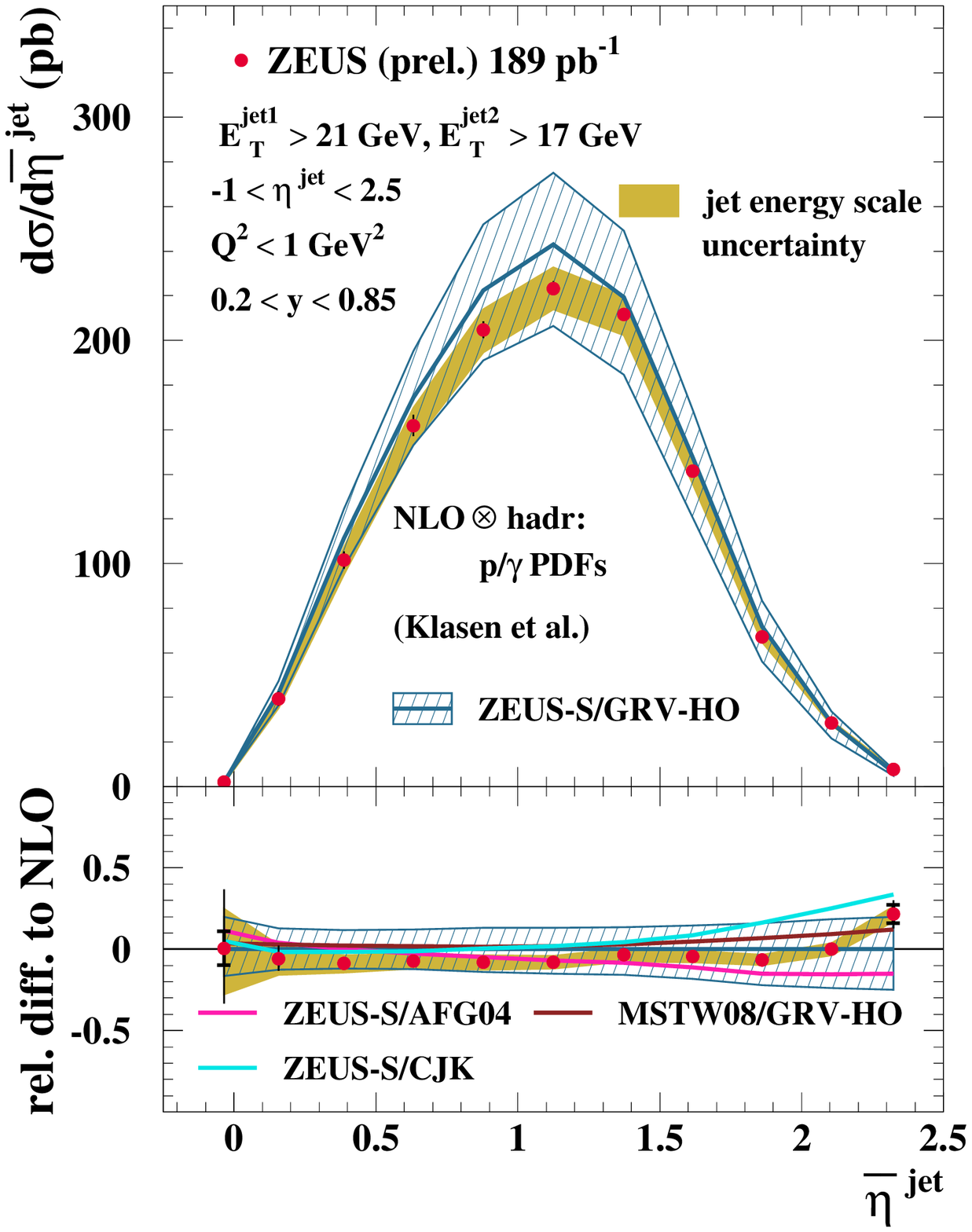}
  \hspace*{-1.0cm}\includegraphics[height = 0.4\textwidth, width=0.39\textwidth, bb = 0 0 550 535, clip = true]{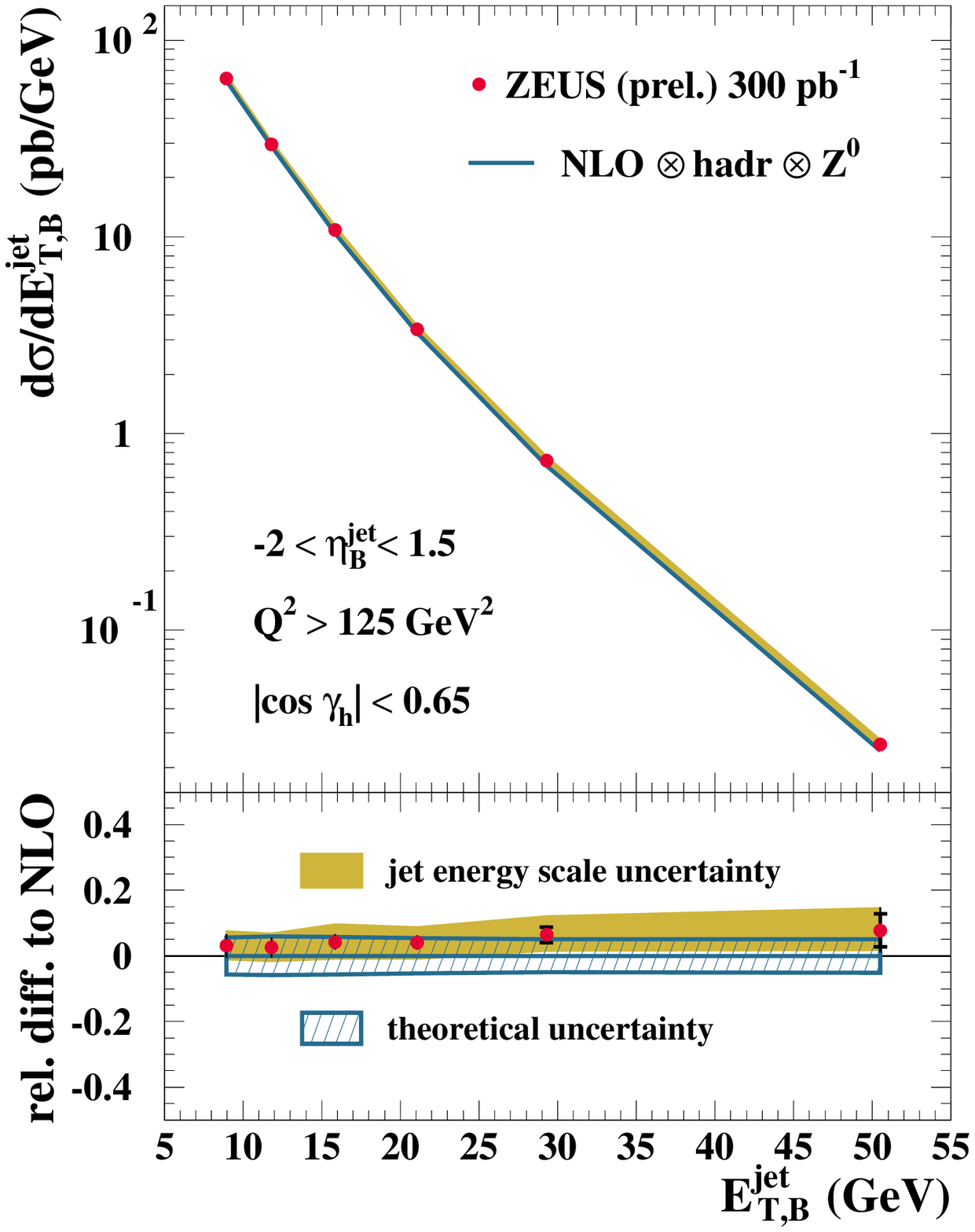}
  \hspace*{-0.8cm}\includegraphics[height=0.31\textwidth,width=0.37\textwidth, bb = 290 365 561 545, clip = true]{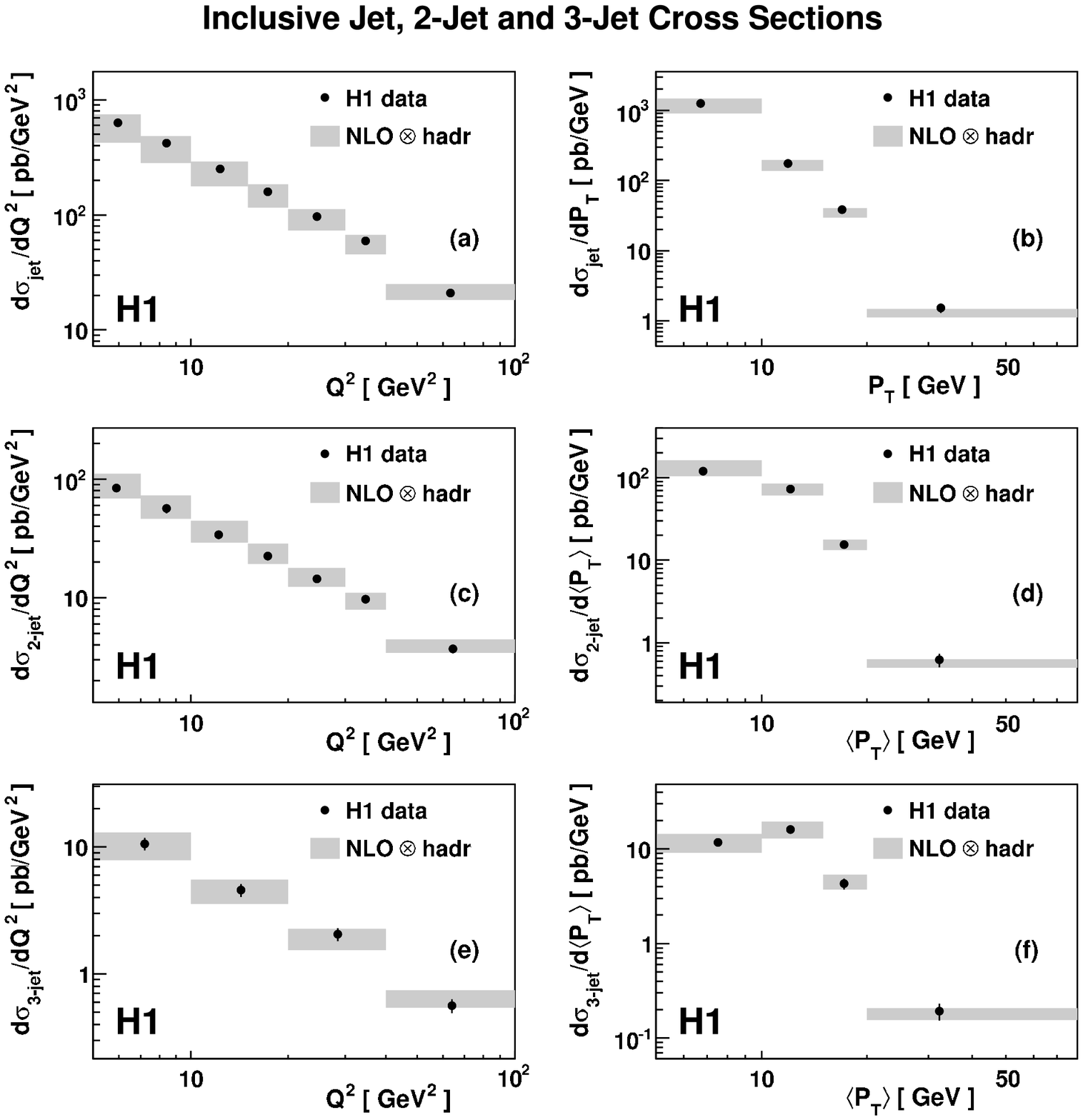}
  \vspace{-10pt}
  {\caption{Dijet cross section as a function of $\bar{\eta}^{\text{jet}}$ in photoproduction~\cite{abr10a} and inclusive-jet cross sections as a function of $E_{\text{T}}^{\text{jet}}$~\cite{abr10c} and $P_{\text{T}}$~\cite{aar10a} in DIS.}
 \label{fig:multijets}}
\end{figure}

Figure~\ref{fig:multijets} shows the dijet cross section as a function
of $\bar{\eta}^{\text{jet}}$ in $\gamma p$ and inclusive-jet cross
sections as a function of $E_{\text{T}}^{\text{jet}}$ and
$P_{\text{T}}$ in DIS. The data are very precise, the dominant
experimental error being the energy scale of the jets in the ZEUS
measurements and the model dependence of the data correction in the H1
measurement.
The measurements are compared to next-to-leading (NLO) QCD predictions
which describe the data very well. The theoretical errors are
dominated by the renormalization scale uncertainty. The dijet cross
sections have the potential of constraining further the proton and
photon PDFs and the inclusive-jet cross sections provide direct
sensitivity to $\alpha_{s}(M_{Z})$.
\subsection{Extraction of $\mathbf{\alpha_{s}}$}
\begin{figure}[h]
    \includegraphics[bb = 0 0 470 510, height = 0.42\textwidth, width=0.44\textwidth]{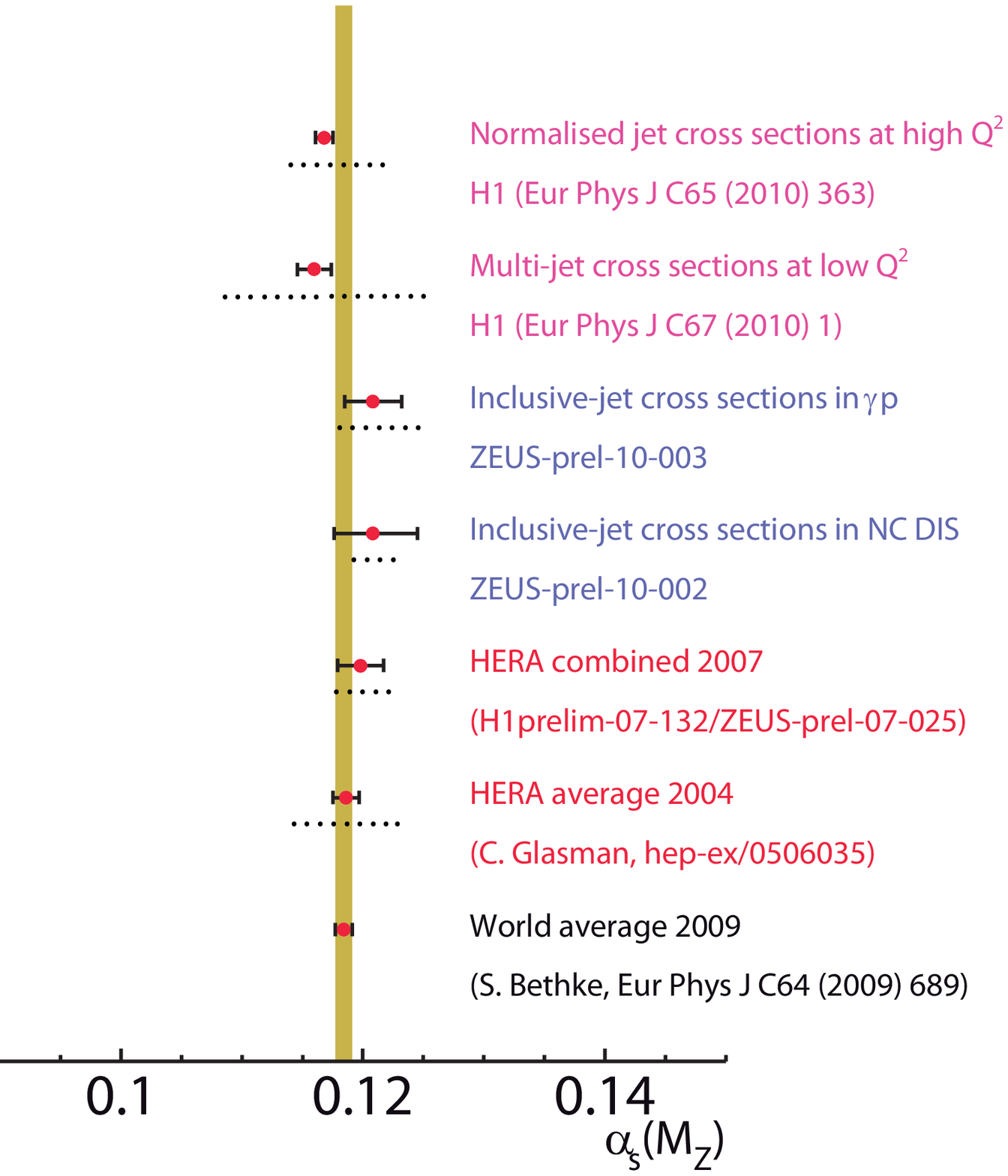}
    \includegraphics[height = 0.38\textwidth,width=0.44\textwidth]{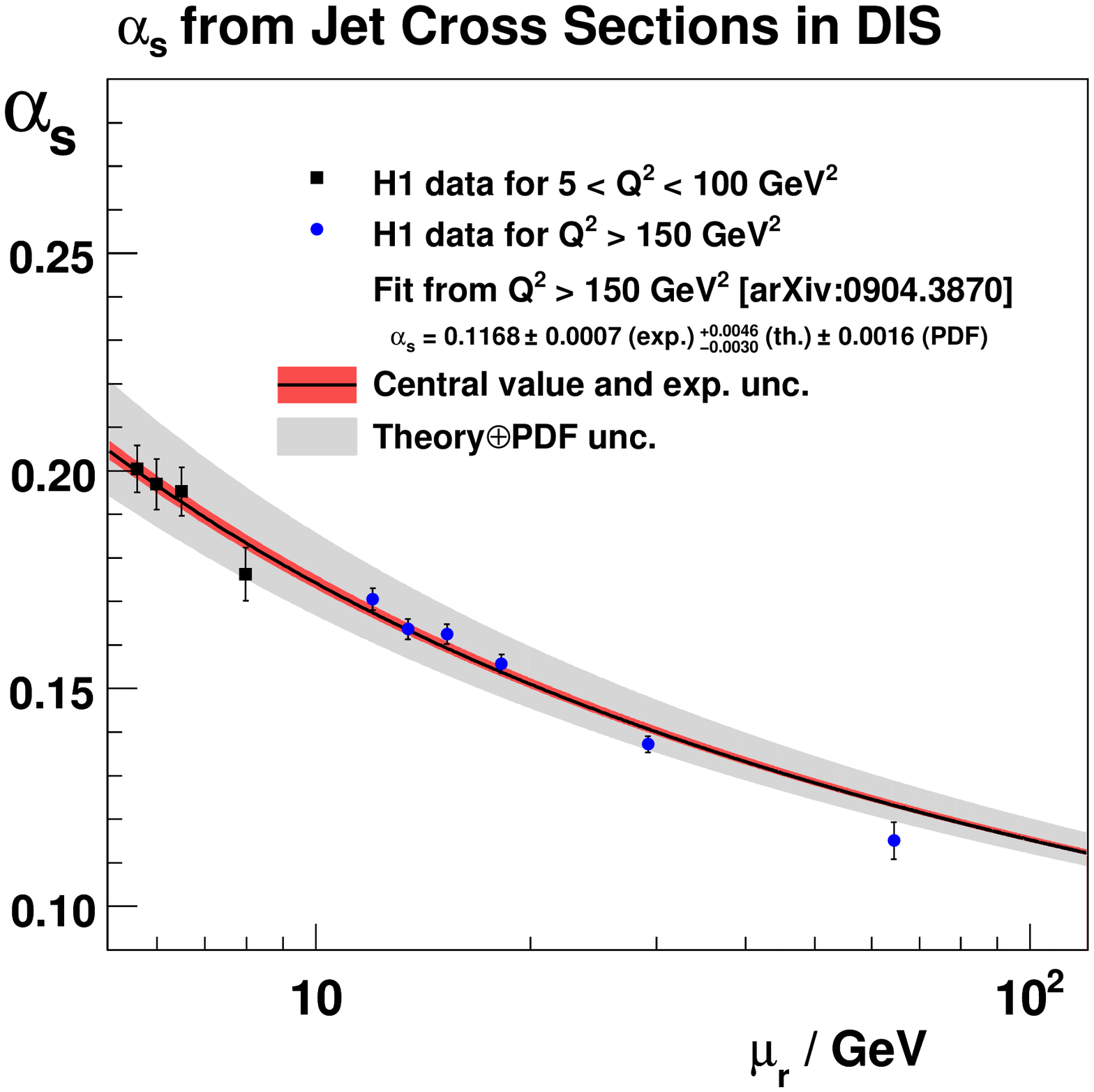}
    \vspace{-10pt} {\caption{Summary of $\alpha_{s}(M_{Z})$ values
        extracted from the H1 and ZEUS data together with the HERA and
        world averages (left). Summary of the running $\alpha_{s}$
        values extracted from the H1 jet data in low and high $Q^{2}$
        (right).}
 \label{fig:alpha_summary}}
\end{figure}

The jet cross sections described above were used to extract the value
of $\alpha_{s}$ and to test its running.
Figure~\ref{fig:alpha_summary} (left) shows a summary of
recent $\alpha_{s}(M_{Z})$ measurements from the H1 and ZEUS
collaborations together with the HERA averages of 2004~\cite{glas05}
and 2007~\cite{h1z07} and the current world average~\cite{bet09}. All
measurements are consistent with each other and with the world
average. For many of the measurements the theoretical uncertainties
dominate the error. Higher order calculations are expected to improve
the results.

The scale dependence of $\alpha_{s}$ was determined by extracting
$\alpha_{s}$ at different values of the scale. Statistical, systematic
and correlated uncertainties were taken into
account. Figure~\ref{fig:alpha_summary} (right) shows the running of
$\alpha_{s}$ as a function of the renormalization scale, where the
values were extracted from the H1 data at low and high
$Q^{2}$. The measurements at low $Q^{2}$ agree well with the QCD
expectations for $\alpha_{s}$ based on the jet cross section
measurements at high $Q^{2}$.
\section{Heavy Flavor Physics at HERA}
In $e^{\pm}p$ collisions, the main production mechanism for heavy
flavors is the boson-gluon fusion process. The large mass of the heavy
quarks produced in this process and large $Q^{2}$ (in case of DIS) or
$p_{\text{T}}$ of the heavy quark (in case of $\gamma p$) provide hard
scales, so that pQCD is applicable. Several different experimental
techniques are used to tag the heavy quark final state. These methods
include identification of a lepton which is produced in the
semileptonic decay of heavy quarks (lepton tag), exploiting the
lifetime information using the impact parameter or decay length
(lifetime tag) and meson identification (e.g.\ $D^{*\pm}$
tagging). Different tagging methods often cover different kinematic
regions and can be combined to enhance the separation between signal
and background.

In a recent ZEUS measurement~\cite{abr10d}, the semileptonic decay of
$b$ hadrons into electrons was combined with lifetime
information. This analysis was performed in the DIS regime ($Q^{2} >
10\xGeV^{2}, 0.9 < p_{\text{T}}^{e} < 8\xGeV{}, |\eta^{e}| <
1.5$). The extraction of the beauty signal was done using a
likelihood-ratio test, which allowed information on electron
identification to be combined with the semileptonic decay
kinematics. The beauty production cross sections were measured in bins
of $Q^{2}, x, p_{\text{T}}^{e}$ and $\eta^{e}$ and the beauty
contribution to the proton structure function was calculated. The
measured cross sections were found to be in good agreement with the
scaled \RAPGAP{} LO MC prediction and NLO QCD predictions calculated
from the HVQDIS program.

The H1 collaboration has recently released a new
measurement~\cite{h110} of $D^{*}$ meson production in DIS ($5 <
Q^{2} < 100\xGeV^{2}, p_{\text{T}}(D^{*}) > 1.25\xGeV{}, |\eta(D^{*})|
< 1.8$). Single and double differential cross sections were determined
and compared to different MC models and NLO QCD predictions. In
general all cross sections were found to be reasonably well described
by different LO and NLO QCD predictions.

Two preliminary results from the ZEUS collaboration on beauty production
in $\gamma p$~\cite{abr10e} ($Q^{2} < 1\xGeV^{2},
p_{\text{T}}^{\text{jet} 1 (2)} > 7 (6)\xGeV{}, -1.6 \le
\eta^{\text{jet} 1 (2)} < 1.3$) and DIS~\cite{abr10f} ($5 < Q^{2} <
1000\xGeV^{2}, -1.6 < \eta^{\text{jet}} < 2.2$), used inclusive
secondary vertices associated to jets.
The beauty content was determined by a $\chi^{2}$ fit of the mirrored
decay length significance distribution in bins of invariant mass of
the secondary vertex.
\begin{wrapfigure}[16]{r}{0.39\textwidth}
 \vspace{-13pt}
   \begin{center}
    \includegraphics[width=0.36\textwidth, clip=true]{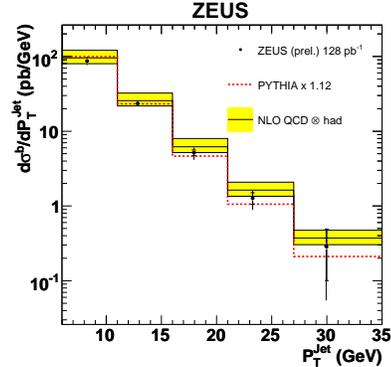}
    \vspace{-10pt} \setlength{\captionmargin}{12pt}
    \caption{Differential cross section for beauty
        production in $\gamma p$ as a function of
        $p_{\text{T}}^{\text{jet}}$, compared with scaled \PYTHIA{}
        and NLO QCD predictions~\cite{abr10e}.}
   \label{fig:php_xsec}
\end{center}
\end{wrapfigure}
The beauty photoproduction cross section as a function of
$p_{\text{T}}^{\text{jet}}$ compared to NLO QCD prediction from FMNR
and the scaled LO \PYTHIA{} is shown in Figure~\ref{fig:php_xsec}. The
measured cross sections are in reasonable agreement with the NLO QCD
prediction and their shape is well described by LO MC prediction.

H1 measurements of cross sections for events with charm and beauty
jets in DIS~\cite{aar10c} ($Q^{2} > 6\xGeV^{2},
E_{\text{T}}^{\text{jet}} > 6\xGeV{}, -1 < \eta^{\text{jet}} < 1.5$)
were also presented. The numbers of charm and beauty jets were
determined using variables reconstructed using the H1 vertex detector.
The measurements were compared with QCD predictions and with previous
measurements. These measurements showed that the production of beauty
and charm jets in DIS is reasonably well described by NLO QCD
predictions.
\subsection{$\mathbf{F_{2}^{b\bar{b}}~ \text{and}~ F_{2}^{c\bar{c}}}$}
\begin{figure}[h]
  \includegraphics[bb = 0 0 420 540,height = 0.5\textwidth,
  width=0.44\textwidth, clip=true]{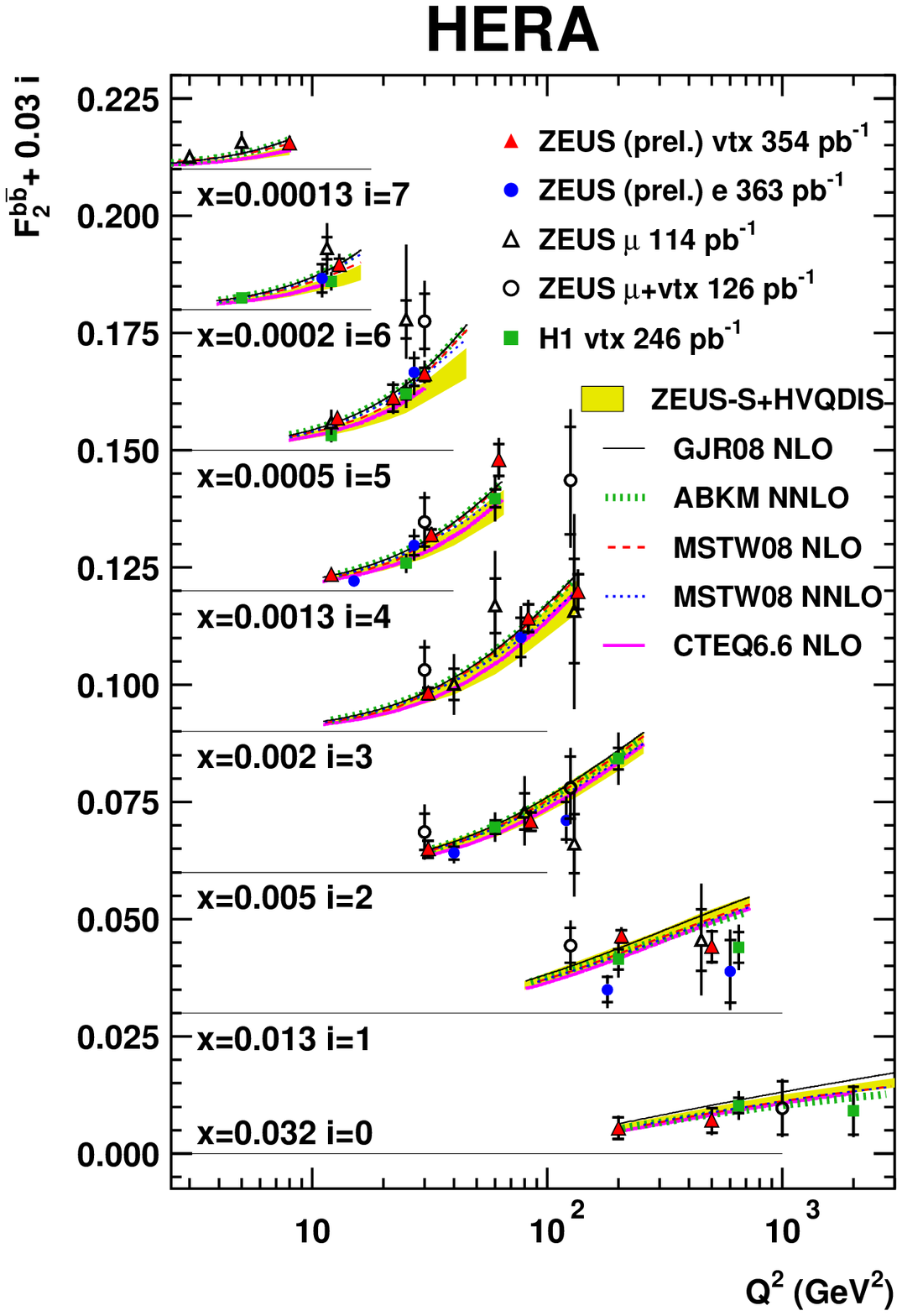}
  \includegraphics[bb = 0 0 704 668, height = 0.48\textwidth,
  width=0.42\textwidth, clip=true]{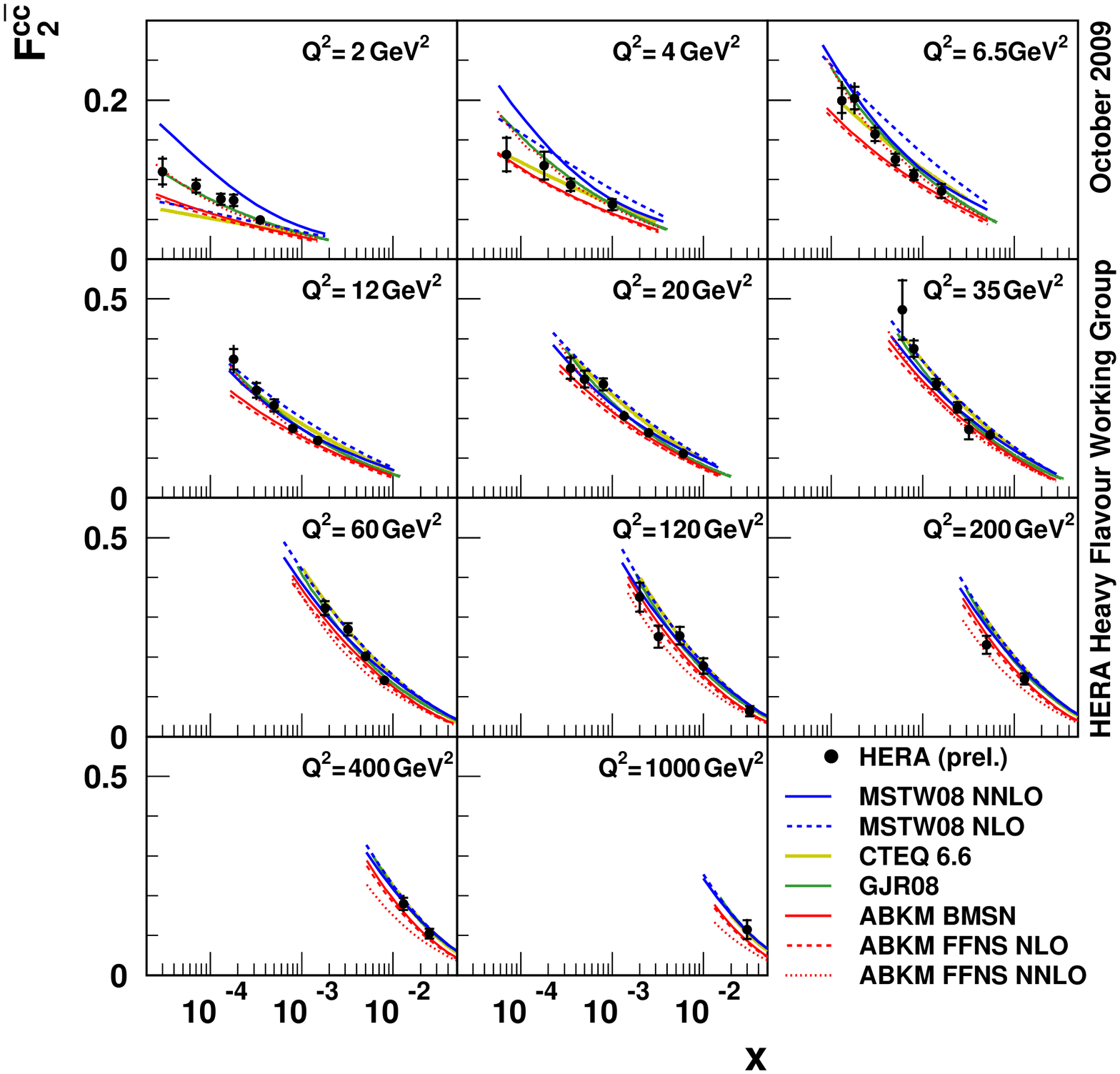}
\vspace{-10pt} {\caption{$F_{2}^{b\bar{b}}$ as a function of $Q^{2}$
    for fixed values of $x$ compared to different theory predictions
    (left). Combined H1 and ZEUS $F_{2}^{c\bar{c}}$ (black dots)
    compared to NLO and NNLO QCD predictions (right).}
  \label{fig:F2b_F2c}}
\end{figure}

The data from beauty production in DIS using the electron decay
channel and inclusive secondary vertexing was used to extract the
beauty contribution to the proton structure function $F_{2}$, denoted
as $F_{2}^{b\bar{b}}$. Figure~\ref{fig:F2b_F2c} (left) shows
$F_{2}^{b\bar{b}}$ as a function of $Q^{2}$ for fixed values of
$x$. The results from two recent ZEUS measurements has been compared
with the previous measurements from H1 and ZEUS. The different
measurements are consistent with each other. Also the results are
compared to several NLO and NNLO QCD predictions. The data are
reasonably well described by the different theory predictions.

The H1 and ZEUS combined $F_{2}^{c\bar{c}}$ was obtained by combining
several different measurements made using different tagging
techniques~\cite{h1z09}. The correlations of the systematic
uncertainties between the different measurements were taken into
account. The data covered the kinematic range of $2 < Q^{2} <
1000\xGeV^{2}$ and $10^{-5} < x < 10^{-1}$. A precision of $5 - 10\%$
was reached for the combined results. The results were compared with
different approaches of pQCD. The comparison with different
predictions (see Figure~\ref{fig:F2b_F2c} (right)) shows that in most
of the $x$--$Q^{2}$ plane the data are more precise than the spread
observed in the theoretical predictions. The $F_{2}^{c\bar{c}}$ data
represents therefore valuable constraints on the theory of heavy
flavor production in DIS.
\section{Summary}
A selection of results of recent measurements for jet and heavy flavor
production in $\gamma p$ and DIS at HERA was presented. Jet physics at
HERA provide high precision QCD measurements. These measurements can
help to constrain further the proton and photon PDFs. Precise and consistent
$\alpha_{s}$ extraction in different kinematic regimes has been
performed. The running of the coupling is verified over a wide range
of the scale.

The measured cross sections for beauty and charm production are in
general consistent with the NLO QCD predictions. The different
measurements provide a consistent picture of $F_{2}^{b\bar{b}}$ and
$F_{2}^{c\bar{c}}$. Combining H1 and ZEUS, $F_{2}^{c\bar{c}}$ results
in precise measurements and provides constraints for theory.


\end{document}